# The Current Ability to Test Theories of Gravity with Black Hole Shadows

Yosuke Mizuno[1], Ziri Younsi[1,2], Christian M. Fromm[1,3], Oliver Porth[1], Mariafelicia De Laurentis[1], Hector Olivares[1], Heino Falcke[4], Michael Kramer[3,5], and Luciano Rezzolla[1,6]

[1]*Institut für Theoretische Physik, Goethe Universität, Max-von-Laue-Str. 1, 60438, Frankfurt am Main, Germany*
[2]*Mullard Space Science Laboratory, University College London, Holmbury St. Mary, Dorking, Surrey, RH5 6NT, UK*
[3]*Max-Planck-Institut für Radioastronomie, Auf dem Hügel 69, Bonn 53121, Germany*
[4]*Department of Astrophysics/IMAPP, Radboud University Nijmegen P.O. Box 9010, 6500 GL Nijmegen, The Netherlands*
[5]*Jodrell Bank Centre for Astrophysics, University of Manchester, M13 9PL, UK.*
[6]*Frankfurt Institute for Advanced Studies, Ruth-Moufang-Str. 1, Frankfurt am Main, 60438, Germany*

**Our Galactic Center, Sagittarius A* (Sgr A*), is believed to harbour a supermassive black hole (BH), as suggested by observations tracking individual orbiting stars[1,2]. Upcoming submillimetre very-long-baseline-interferometry (VLBI) images of Sgr A* carried out by the Event-Horizon-Telescope Collaboration (EHTC)[3,4] are expected to provide critical evidence for the existence of this supermassive BH[5,6]. We assess our present ability to use EHTC images to determine if they correspond to a Kerr BH as predicted by Einstein's theory of general relativity (GR) or to a BH in alternative theories of gravity. To this end, we perform general-relativistic magnetohydrodynamical (GRMHD) simulations and use general-relativistic radiative transfer (GRRT) calculations to generate synthetic shadow images of a magnetised accretion flow onto a Kerr BH. In addition, and for the first time, we perform GRMHD simulations and GRRT calculations for a dilaton BH, which we take as a representative solution of an alternative theory of gravity. Adopting the VLBI configuration from the 2017 EHTC campaign, we find that it could be extremely difficult to distinguish between BHs from different theories of gravity, thus highlighting that great caution is needed when interpreting BH images as tests of GR.**

The defining feature of a BH is the existence of an "event horizon", a region of spacetime within which neither matter nor radiation can escape. Just outside of the event horizon is a region wherein photons follow unstable orbits. The precise size and shape of this "photon region" depends on the properties of the BH[7,8]. Therefore, when a BH is observed directly, we expect to see its "shadow" as a manifestation of this photon region on the background sky[9]. Although astrophysical BHs are expected to be described by the Kerr solution in GR, numerous BH solutions exist in other theories of gravity.

To obtain a realistic BH shadow image, it is important to understand both the plasma dynamics (accretion and outflow) around the central BH and its radiative properties. Calculations of BH shadow images from GRRT and GRMHD simulations of BH accretion have been performed by several groups[10–14]. However, in all of these simulations only Kerr BHs were considered. Therefore, to utilise the future results of the EHTC as a test of GR, it is crucial to investigate the plasma dynamics around BHs different from the Kerr solution and study their subsequent radiative properties. This is the rationale behind our work, where we have performed, for the first time, GRMHD simulations and GRRT calculations of accretion flows onto BHs that are not general-relativistic solutions and are based on the Rezzolla-Zhidenko parameterisation of BH solutions in general metric theories of gravity[15]. The results of these calculations are used to generate synthetic BH shadow images considering a VLBI configuration similar to that of the 2017 EHTC observational campaign.

The first BH solution considered is a *rotating* Kerr BH with dimensionless spin parameter $a_* := J/M^2 = 0.6$, with $J$ and $M$ the BH's angular momentum and mass, respectively. This value is chosen because current comparisons of the accretion flow with broadband spectra and past EHTC observations via semi-analytic models yield a rather low $a_*$ value[16,17], while GRMHD models have so far explored comparatively high values of $a_*$[11,12]. The second BH solution considered is an asymptotically flat dilaton BH[18]. This BH appears as a solution in Einstein-Maxwell-dilaton-axion gravity, where the Einstein-Maxwell equations are coupled to the scalar dilaton and axion fields. Our choice for a dilaton BH as a representative case of an alternative theory of gravity is motivated by the fact that such a BH solution is physically distinct from GR since the action now introduces a fundamental (dilaton) scalar field representing the low-energy regime of the bosonic sector in heterotic string theory[18].

To make the comparison between the two BHs even more striking, we consider the dilaton BH to be *non-rotating* by assuming both the axion field and BH spin vanish. The only degree of freedom is then given by the dimensionless "dilaton parameter", $\hat{b}_* := \hat{b}/M$, where $\hat{b}$ is the "dilaton parameter". Typically[19], $0 \leq \hat{b}_* \leq 1$ and $\hat{b}_*$ may be set so that either the event horizon, the unstable circular photon orbit, or the innermost stable circular orbit (ISCO) coincide for both BHs [see Supplementary Information (*SI*) for details]. Out of these three options, we first consider equating the ISCOs of the two BHs, which occurs for $\hat{b}_* = 0.504$. This choice enforces similarities in the dynamics of the accretion flow at the inner edge of the accretion disc. However, the other two options have also been considered, yielding similar results that are presented in the *SI*.

We start our state-of-the-art GRMHD three-dimensional (3D) simulations from a torus in hydrodynamic equilibrium with a weak poloidal magnetic field, using the code BHAC[14]. During the evolution, the magnetic field is triggered to excite the magnetorotational instability (MRI) in the torus and a turbulent structure develops[14,20]. Plasma in the torus then accretes onto the central BH, but also outflows from the torus as a "disc wind". The region near the polar axis, termed "funnel", has a low rest-mass density and is highly magnetised with a mostly ordered magnetic field. The region above the turbulent accretion torus, termed "disc corona", is of moderately low density and low magnetisation, with a tangled magnetic field. In between the magnetised funnel and the disc-wind regions a strong shear develops, which has been identified with a jet[21]. This is shown in Fig. 1, where we report the density $\rho$ (left panel) and magnetisation $\sigma := b^2/\rho$ (right panel), where $b$ is the magnetic-field strength in the comoving frame. In both panels, the left half refers to the *non-rotating* dilaton BH, while the right half to the *rotating* Kerr BH. The data is time-averaged over the interval $t = 11000 - 12000\,M$, which is when the simulations have reached a quasi-steady state and turbulence in the plasma is fully developed in the inner regions of the flow, and also corresponds to the typical timescale of VLBI observations (i.e., $\sim 6\,\mathrm{hours}$). When comparing the two cases, including quantifications of mass-accretion rate and magnetic flux through the horizon (see *SI*), it is apparent that the overall structures are very similar, so that, from a plasma-dynamics viewpoint, the dilaton BH closely mimics the Kerr BH.

To model the horizon-scale emission satisfying present observational constraints of Sgr A*, we perform GRRT calculations of the GRMHD simulation data using the code BHOSS (Younsi et al. 2018 in prep.). Figure 2 shows the median thermal synchrotron emission image at $230\,\mathrm{GHz}$ for an accreting BH having the same mass, distance, and accretion rate associated with Sgr A*. The left panel refers to the Kerr BH and the middle panel to the dilaton BH; in both cases the images are averaged over an interval of $\sim 6\,\mathrm{hours}$; variability on smaller timescales will only decrease the quality of the image and we neglect it here to consider a "best-case" scenario. Clearly, there are visual differences between the two images such as: the distribution of the intensity across the approaching front limb of the torus, the intensity near the horizon, and the diameter and centre of the shadow. For instance, from the pixel-by-pixel difference (right panel in Fig. 2), it is clear that the dilaton BH shadow is smaller than the shadow of the Kerr BH because of the smaller horizon of the dilaton BH. Also visible are the offsets between the two shadows (red ring) and the greater asymmetry in the brightness profile of the shadow for the Kerr BH, which is due to its intrinsic spin[22].

Using Fig. 2 it is tempting to draw conclusions about our ability to discriminate between the images of a Kerr and of a dilaton BH. However, the comparison presented thus far refers to "infinite-resolution images" from GRRT calculations. To properly assess this ability one must take into account the realistic properties of the VLBI array and stations performing the observations, together with the various systematic effects, both physical and computational, that contaminate the final image. We recall that VLBI observations provide an incomplete sample of visibilities (amplitude and phase), which are the Fourier transforms of the flux-density distribution of the source. Reconstruction algorithms are needed to obtain an image from the VLBI observations and we here use the Bi-Spectrum Maximum Entropy Method (BSMEM)[23]; however, to confirm our results we additionally employ the linear polarimetric MEM (PolMEM)[24] (see *SI*). The simulated visibilities of synthetic observation of Sgr A* are obtained from the EHT imaging software[24], adjusted to the realistic conditions of 2017 EHTC observations (see *SI*). Indeed, already a non-imaging inspection of the synthetic dataset shows that the variation of the closure phases, i.e., of the sum of the phases of a closed antenna triangle, presents non-zero closure phases, thus indicating an asymmetric source structure (see *SI*).

For the synthetic images we use a scan length of $420\,\text{s}$, an integration time of $12\,\text{s}$, and include interstellar scattering. The reconstructed image is convolved with $50\%$ of the nominal beam size, which is shown on the bottom left corner of each panel. In Fig. 3 we report the convolved GRRT images (left panels), the reconstructed images using BSMEM without the effect of interstellar scattering (middle panels), and the reconstructed images using BSMEM and including interstellar scattering (right panels), both in the case of a Kerr BH (upper row) and of a dilaton BH (lower row). Clearly, although we have used only $50\%$ of the nominal beam size, the convolved images already have smeared out the sharp emission features seen in the original GRRT images, while the reconstructed images using BSMEM have mapped critical features of the BHs images into noisy features. The image sharpness further deteriorates with the inclusion of interstellar scattering, which increases the blurring of these features and obviously lowers the signal-to-noise ratio.

For a more quantitative comparison we compute two image-comparison metrics: the mean square error (MSE) and the structural dissimilarity (DSSIM) index[25], which we report in Table 1, noting that for both indices, zero values correspond to a perfect match between images. Our image-comparison procedure consists of two steps to ensure that we compare images of identical resolution. First, we generate the "convolved" image of the ray-traced snapshot using the observing beam as constrained by the VLBI array. Second, we produce the "reconstructed" image after convolving the BSMEM algorithm with the observing beam. Overall, Table 1 indicates that for both BHs the prominent features of the original convolved images are well-matched in the reconstructed images (see first two rows of Table 1). However, similar matches are obtained when comparing the convolved image of a Kerr BH with the reconstructed image of a dilaton BH, and vice-versa (see last two rows of Table 1). Our results demonstrate, using non-negligible values of the dilaton parameter, that it is still difficult to use upcoming VLBI images of Sgr A* at 230 GHz to unambiguously distinguish between two mathematically and physically distinct BH spacetimes. In turn, these results cast doubt on our present ability to test GR by making use of such images, suggesting that great caution is needed when utilising BH images as a test of GR.

The results we have presented focus on a specific example of a non-general-relativistic BH solution and do not consider the case of extremal black holes. They demonstrate that it is presently difficult to distinguish between a Kerr BH and a dilaton BH on the basis of BH shadow images alone, and hence to measure the spin of the black hole. This suggests that for a given alternative, non-general-relativistic BH solution, it would still prove difficult to distinguish between GR and non-GR spacetimes. These results motivate employing an approach independent of any specific theory of gravity, and through which a parametric representation of the BH spacetime is used, rather than laboriously checking every possible theory in turn.

In conclusion, we note that several future developments can improve our ability to discriminate between GR and alternative theories of gravity using shadow images. These include more advanced image reconstruction algorithms, future increases in observational frequency (e.g., $340\,\mathrm{GHz}$; see *SI*) and bandwidth (e.g., $8\,\mathrm{GHz}$), the inclusion of additional VLBI antennas (e.g., in Africa; see *SI*), source variability and timing measurements, and concurrent multifrequency spectroscopic and spectro-polarimetric observations together with horizon-scale VLBI shadow images. Pulsar timing[26] observations in the vicinity of Sgr A* also have the potential to impose stringent constraints on the underlying theory of gravity[27].

**Methods**

In what follows we describe the methods and assumptions employed to derive our results.

**General-Relativistic Magnetohydrodynamic Simulations.** The GRMHD simulations were performed using the recently developed `BHAC` code[14], which solves the equations of GRMHD in arbitrary but fixed spacetimes and employs a number of different coordinate systems; the results presented here refer to 3D simulations using spherical polar coordinates. As initial data we adopt a standard setup in simulations of accretion flows onto BHs, namely a torus with a single weak poloidal magnetic field loop in hydrodynamic equilibrium around a central BH. Following the setup discussed by Meliani et al. (2017)[28], the equilibrium torus model is chosen to have a constant distribution of specific angular momentum, which in turn determines the location of the inner edge of the torus $r_{\rm in}$[29,30]. In the case of the Kerr BH we choose $\ell_0/M = 4.5$ so that $r_{\rm in} = 10.3\,M$, while for the dilaton BH we set $\ell_0/M = 4.567$ and thus $r_{\rm in} = 10.3\,M$.

Whilst an analytic form for the metric of a general dilaton BH is known, and hence also for the non-rotating case, we here adopt its representation via the general parameterisation of BH spacetimes in arbitrary metric theories of gravity. This approach was recently developed by Rezzolla and Zhidenko for spherically symmetric spacetimes[15] and more recently extended to axisymmetric spacetimes[31] [see also Johannsen and Psaltis (2011)[32] for an approach aimed at specific modifications of the Kerr metric]. In practice, this method employs an efficient mathematical expansion of BH metrics in terms of a rapidly converging series through which any BH metric can be recast (see *SI*). The advantage of this approach is that it provides a general and agnostic framework in which to perform the analysis of accretion flows onto BHs or the dynamics of test objects on orbits close to the BH[33]. More specifically, in the case of a non-rotating dilaton BH we use the expansion coefficients $\epsilon$, $a_0$, $b_0$, $a_1$, $b_1$, $a_2$ and $b_2$, which yield a sub-per-mill relative difference between the exact and the approximated metric for the metric functions. The original expanded metric has a coordinate singularity at the BH event horizon, which we remove by rewriting the original spherically-symmetric metric in horizon-penetrating coordinates.

The GRMHD simulations make use of an ideal-gas equation of state[30] with an adiabatic index of $\Gamma = 4/3$, while the poloidal magnetic field loop is defined in terms of the vector potential: $A_\phi \propto \max(\rho/\rho_{\max} - 0.2, 0)$ and normalised so that $\beta_{\min} = (2p/b^2)_{\min} = 100$, where the subscript min refers to the minimum value inside the torus. As is customary in this type of simulation, we excite the MRI in the torus by adding a $1\%$ random perturbation to the equilibrium gas pressure of the torus. To avoid the presence of vacuum regions outside the torus it is conventional to add an "atmosphere" to regions which should be virtually in vacuum. In essence, floor values are applied for the density ($\rho_{\mathrm{fl}} = 10^{-4}\, r^{-3/2}$) and the gas pressure ($p_{\mathrm{fl}} = (10^{-6}/3)\, r^{-5/2}$). In practice, for all numerical cells which satisfy $\rho \leq \rho_{\mathrm{fl}}$ or $p \leq p_{\mathrm{fl}}$, we simply set $\rho = \rho_{\mathrm{fl}}$ and $p = p_{\mathrm{fl}}$; no prescription is imposed on the fluid velocity.

The spherical polar coordinate map $(r, \theta, \phi)$ covers the simulation domain in the range $r \in [0.8\, r_{\mathrm{h}}, 1000\, M]$, $\theta \in [0.01\pi, 0.99\pi]$, and $\phi \in [0, 2\pi]$, where $r_{\mathrm{h}}$ is the event-horizon radius, and with the innermost cell well inside the event horizon. The grid spacing is logarithmic in radial and uniform in $\theta-$ and $\phi-$ directions, with the number of gridpoints given by $(N_r, N_\theta, N_\phi) = (256, 128, 128)$. At the inner/outer radial boundaries we apply standard inflow/outflow boundary conditions by enforcing a copy of the physical variables, while at the polar boundaries hard polar boundary conditions proposed by Shiokawa et al. (2012)[34] are applied: a solid reflective wall along the polar boundaries where the flux through the boundaries is set to be zero manually, adjusting the electromotive force in the constrained transport routine, and setting the poloidal velocity along the polar boundaries to zero. For the azimuthal direction, instead, periodic boundary conditions are employed.

**General-Relativistic Radiation-Transfer Calculations.** In order to generate images for EHTC observations of Sgr A*, we perform GRRT of `BHAC`'s GRMHD simulations using the `BHOSS` code (Younsi et al. 2018 in prep.). In this code, electromagnetic radiation propagates along null geodesics of the spacetime and we solve both the geodesic equations of motion and the radiative-transfer equation in tandem[35]. In practical modelling of radiative emissions from accretion flows, a statistically stationary flow is desirable. In practice, however, the total mass within the torus is finite and accreted onto the BH over time.

To produce the actual images, we take the time interval in all cases as $11000 - 12000\,M$, which is $\sim 6\,\text{hours}$ given the estimated mass and distance of Sgr A*, which we take to be $4.02 \times 10^6\,\text{M}_\odot$ and $7.86\,\text{kpc}$, respectively [36]. Other physical assumptions are a thermal distribution function for the radiating electrons (which emit Synchrotron radiation and are also self-absorbed), and in the actual construction of the image the fixed ion-to-electron temperature ratio[37], $T_\text{i}/T_\text{e}$, the mass-accretion rate, $\dot{M}$, and the observer inclination angle, which we fix at $60°$. The latter assumption is supported by the small emitting region of the $230\,\text{GHz}$ image recorded in recent VLBI observations of Sgr A*[38]. As a result of these assumptions, the only free parameters in our modelling are $T_\text{i}/T_\text{e}$ and $\dot{M}$, which we normalise at a resolution of $1024 \times 1024$ pixels for both simulations to reproduce Sgr A*'s observed flux at $230\,\text{GHz}$ of $\simeq 3.4\,\text{Jy}$[39]. More specifically, we set $T_\text{i}/T_\text{e} = 3$ for both the Kerr BH and the dilaton BH, and set $\dot{M} = 2.8 \times 10^{-9}\,M_\odot\,\text{yr}^{-1}$ for the Kerr BH and $\dot{M} = 4.3 \times 10^{-9}\,M_\odot\,\text{yr}^{-1}$ for the dilaton BH.

**Synthetic Imaging.** In order to speed up and reduce the computational costs of the imaging analysis, we resample the ray-traced BH shadow images seen at 230 GHz. Furthermore, to avoid aliasing effects, before rescaling the image we convolve the initial ray-traced images with a circular Gaussian having a convolution kernel with full width at half maximum (FWHM) equal to the resolution of the rescaled image. As a result, the rescaled convolved images are $256 \times 256$ pixels in size, with a pixel scale of $0.83\,\mu\text{as}$, and do not lose any intrinsic features from the original GRRT images (cf. left and middle panels in Fig. 3).

For the computation of the synthetic images we use eight antennas scattered across North- and South-America, Europe and the South Pole, which are matched with a possible configuration of the April-2017 EHTC observations. More specifically, they are: the James Clerk Maxwell Telescope (JCMT), the Submillimeter Array (SMA), the Arizona Radio Observatory Submillimeter Telescope (SMT), the Large Millimeter Telescope (LMT), the phased Atacama Large Millimeter/submillimeter Array (ALMA), the Institut de Radioastronomie Millimetrique (IRAM) 30-meter telescope on Pico Veleta (PV), the IRAM Plateau de Bure (PDB), and the South Pole Telescope (SPT). In Supplementary Table 1 we provide an overview of the typical antenna properties, such as the effective antenna diameter $d$ and the system equivalent flux density (SEFD). Our chosen SEFD includes typical atmospheric contributions for the observation of Sgr A* and $10\%$ phasing losses for phased arrays[40,41].

To make our simulations as realistic as possible and provide a test base for the upcoming results of the April-2017 EHTC observation campaign, we assume a total on-source time of $6\,\text{hours}$, from $8{:}30\,\text{UCT}$ until $14{:}30\,\text{UCT}$. The observations of Sgr A* are performed in a typical switching experiment consisting of on-source measurements and off-source calibration and pointing. We list average values obtained from the 2017 EHTC schedule in Supplementary Table 2. Note that no European telescopes, i.e., PV and PDB, participated in our synthetic observations. For the simulations of the synthetic observations of Sgr A*, the EHTC imaging package[24] is used to obtain simulated visibilities. The visibilities generated by the EHTC imaging package are normalised and then converted to the OIFITS format for processing in image reconstruction codes. Finally, the coverage of the $u$-$v$ plane for the April-2017 EHTC observation is presented in Supplementary Fig. 4.

Several algorithms are available for the reconstruction of sparsely-sampled VLBI data and one of the most celebrated is the classical CLEAN algorithm[42]. However, recent work by Chael et al. (2016)[24] has shown that maximum-entropy methods (MEM) are superior in the case of sparse and heterogeneous arrays such as those within the EHTC, justifying the use of BSMEM[23] for the reconstruction of our synthetic images. In order to reproduce conditions that are as realistic as possible, thermal noise and the random errors on the telescope gains, drawn from a Gaussian distribution with unit mean and standard deviation of 0.1, are included during the creation of the visibilities [see Supplementary Fig. 5 and Chael et al. (2016)[24]]. The baseline-dependent thermal noise is calculated from the integration time, $\tau$, the SEFD of the telescope pair and the bandwidth, $\Delta\nu$, according to

$$\sigma_{\text{th},ij} := \frac{1}{0.88}\sqrt{\frac{\text{SEFD}_i\,\text{SEFD}_j}{\tau\Delta\nu}}\,, \qquad (1)$$

where $i$ and $j$ refer to the telescopes in the baseline.

**Image-Fidelity Assessment.** Finally, to quantify the quality of the reconstructed BH-shadow images we use two image quality metrics: the mean square error (MSE) and structured dissimilarity (DSSIM) index. The MSE is a pixel-by-pixel comparison metric that is calculated by averaging the squared intensity difference between two image pixels, namely

$$\text{MSE} := \frac{\sum_{j=1}^{N}|I_j - K_j|^2}{\sum_{j=1}^{N}|I_j|^2}\,, \qquad (2)$$

where $I_j$ and $K_j$ are the $j$-th pixels of the images $I$ and $K$, each with $N$ pixels. The DSSIM is then computed in terms of the human visual-perception metric, i.e., via the so-called "structural similarity" index (SSIM), so that $\text{DSSIM} := 1/|\text{SSIM}| - 1$[43]. Given a pair of images referred to as $I, K$, the SSIM can be expressed as the product of the "luminance" $\mathcal{L}(I, K)$, with the "structure" $\mathcal{S}(I, K)$, and the "contrast", $\mathcal{C}(I, K)$, namely

$$\text{SSIM}(I, K) := \mathcal{L}(I, K)\,\mathcal{S}(I, K)\,\mathcal{C}(I, K)\,. \qquad (3)$$

The individual terms in the equation above depend on the average value of each image, $\mu_{I,K}$, the variance $\sigma_{I,K}^2$ of the individual image, and the covariance of the image pair, $\sigma_{IK}$. Thus, Eq. (3)

can be rewritten as

$$\text{SSIM}(I, K) := \left(\frac{2\mu_I \mu_K}{\mu_I^2 + \mu_K^2}\right) \left(\frac{2\sigma_I \sigma_K}{\sigma_I^2 + \sigma_K^2}\right) \left(\frac{\sigma_{IK}}{\sigma_I \sigma_K}\right) = \left(\frac{2\mu_I \mu_K}{\mu_I^2 + \mu_K^2}\right) \left(\frac{2\sigma_{IK}}{\sigma_I^2 + \sigma_K^2}\right), \quad (4)$$

with

$$\mu_I := \sum_{i=1}^{N} \frac{I_i}{N}, \tag{5}$$

$$\sigma_I^2 := \frac{\sum_{j=1}^{N}(I_j - \mu_j)^2}{(N-1)}, \tag{6}$$

$$\sigma_{IK} := \frac{\sum_{j=1}^{N}(I_j - \mu_I)(K_j - \mu_K)}{(N-1)}. \tag{7}$$

Using these definitions, two identical images would have $\text{MSE} = 0 = \text{DSSIM}$ and $\text{SSIM} = 1$. Before calculating the indices we perform a 2D cross-correlation of the convolved images, i.e., registering the images using common features.

**Correspondence** Correspondence and requests for materials should be addressed to Yosuke Mizuno. (email: mizuno@th.physik.uni-frankfurt.de).

**Acknowledgements** It is pleasure to thank Monika Moscibrodzka, Thomas Bronzwaer, and Jordy Davelaar for fruitful discussions. This research is supported by the ERC synergy grant "BlackHoleCam: Imaging the Event Horizon of Black Holes" (Grant No. 610058). ZY acknowledges support from an Alexander von Humboldt Fellowship. HO is supported in part by a CONACYT-DAAD scholarship. The simulations were performed on LOEWE at the CSC-Frankfurt and Iboga at ITP Frankfurt.

**Author Contributions**   YM performed the GRMHD simulations and wrote the manuscript. ZY performed the GRRT calculations and authored the GRRT code `BHOSS`. CF calculated the synthetic BH shadow images. ZY and CF helped write the manuscript. OP authored the GRMHD code `BHAC`. MDL provided information on alternative theories of gravity. HO helped to perform the GRMHD simulations. HF and MK provided insight into the scientific interpretation of the results. LR initiated the project, closely supervised it and wrote the manuscript. All authors discussed the results and commented on all versions of the manuscript.

**Data availability statement**   The data that support the plots within this paper and other findings of this study are available from the corresponding author upon reasonable request.

**Competing Interests**   The authors declare that they have no competing financial interests.

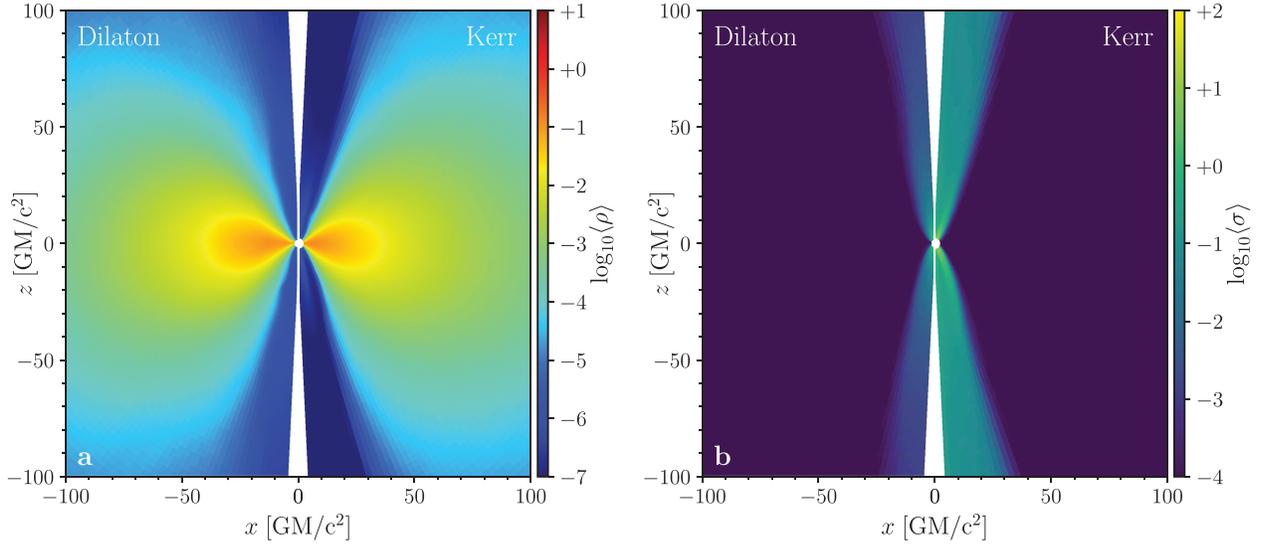

Figure 1: **GRMHD simulations of a magnetised torus accreting onto a Kerr BH and onto a non-rotating dilaton BH.** Panels **a** and **b** show respectively the azimuthal and time-averaged rest-mass density $\rho$ and the magnetisation $\sigma = b^2/\rho$ for a non-rotating dilaton BH with $\hat{b}_* = 0.504$ (*left side of the panels*) and a Kerr BH with $a_* = 0.6$ (*right side of the panels*). The averaging is performed over the time interval $t = 11000 - 12000\,M$, which is when the simulations have reached a quasi-steady state and is also the typical timescale of VLBI observations (i.e., $\sim 6\,\mathrm{hours}$).

Table 1: Image statistics. The mean square error (MSE) and the structural dissimilarity (DSSIM) index between Image 1 and Image 2. Italic letters *a–f* correspond to the various panels in Fig. 3, while the values obtained when Image 2 also includes interstellar scattering are indicated in square brackets. Small values of the MSE or the DSSIM represent very well-matched images.

| Image 1 | Image 2 | MSE$_{1,2}$ | DSSIM$_{1,2}$ |
|---|---|---|---|
| Kerr *a* | Kerr *b* [Kerr *c*] | 0.016 [0.016] | 0.31 [0.31] |
| Dilaton *d* | Dilaton *e* [Dilaton *f*] | 0.016 [0.016] | 0.40 [0.35] |
| Kerr *a* | Dilaton *e* [Dilaton *f*] | 0.018 [0.015] | 0.33 [0.30] |
| Dilaton *d* | Kerr *b* [Kerr *c*] | 0.016 [0.016] | 0.37 [0.37] |

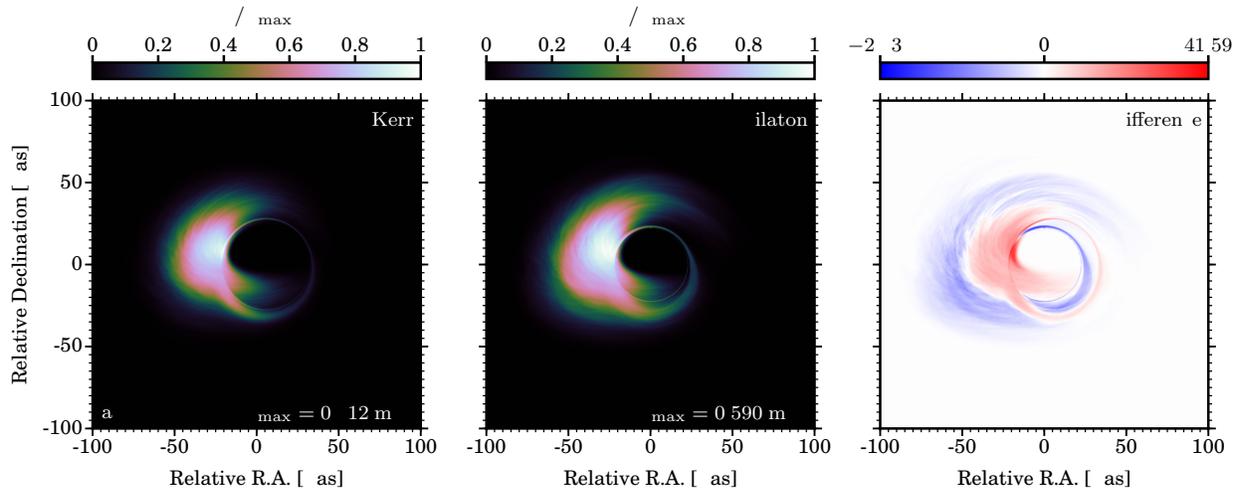

Figure 2: **Simulated BH shadow images of Sgr A* from GRMHD simulations of an accretion flow onto a BH.** Panel **a**: six-hour-averaged BH shadow image of Sgr A* from GRMHD simulations of an accretion flow onto a Kerr BH. Panel **b**: the same as in panel **a**, but for a non-rotating dilaton BH. Panel **c**: Pixel-by-pixel image difference of panels **a** and **b**. The colour scale is linear with red marking pixels for which the Kerr BH image is brighter, while blue pixels indicate the dilaton image is brighter.

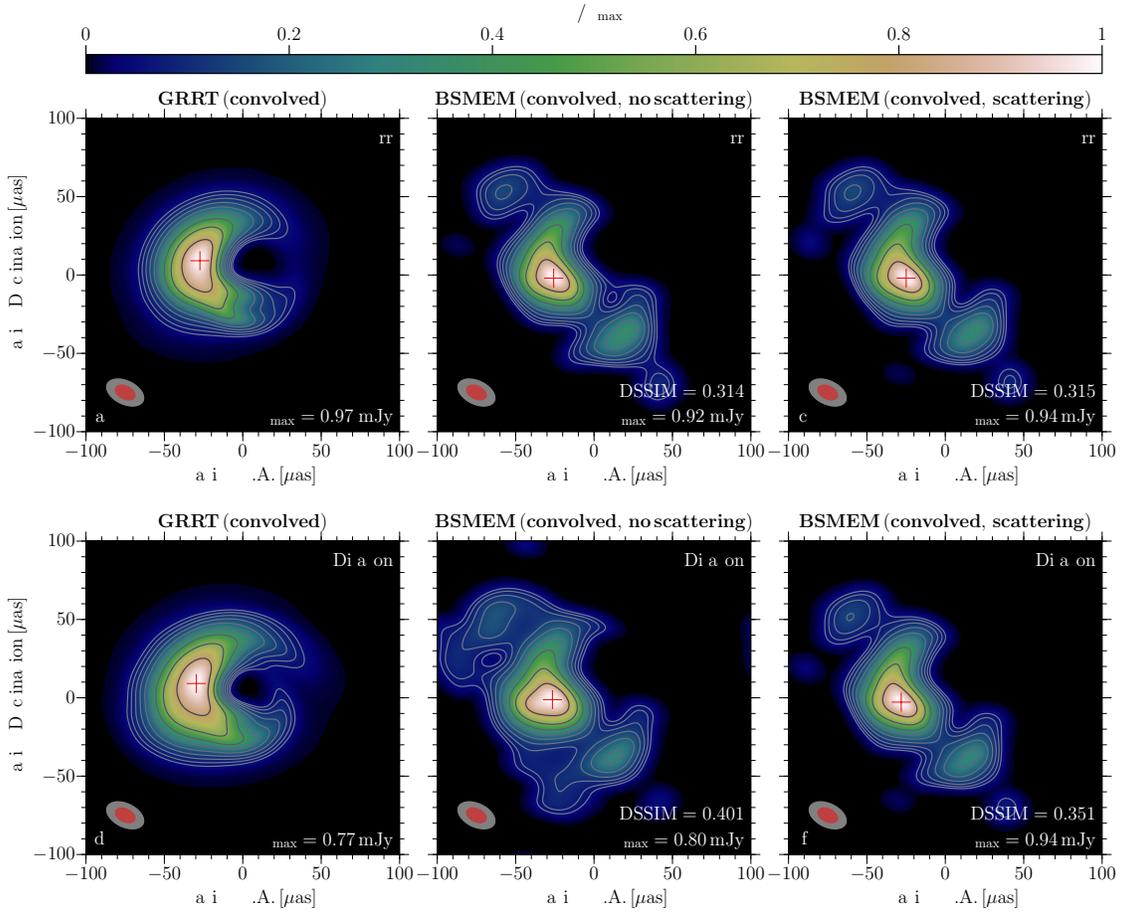

Figure 3: **Synthetic shadow images of Sgr A\* for a Kerr BH and a non-rotating dilaton BH.** From left to right and for both rows: GRRT image convolved with 50% (red shading) of the nominal beam size (light gray shading), the contour levels start at 5% of the peak value and increase by $\sqrt{2}$. The red cross in the images marks the position of the flux density maximum. Panels **a** and **d**: reconstructed image without interstellar scattering, panels **b** and **e**: using BSMEM convolved with 50% (red shading) of the nominal beam size (light gray shading, panels **c** and **f**: reconstructed image including the effect of interstellar scattering using BSMEM. Both images are based on visibilities which take into account a possible VLBI antenna configurations and schedules of EHTC April-2017 observations. The convolving beam size is plotted in the lower left corner of each panel (see *SI*).

# Supplementary Information: The Current Ability to Test Theories of Gravity with Black Hole Shadows

Yosuke Mizuno[1], Ziri Younsi[1,2], Christian M. Fromm[1,3], Oliver Porth[1], Mariafelicia De Laurentis[1], Hector Olivares[1], Heino Falcke[4], Michael Kramer[3,5], and Luciano Rezzolla[1,6]

[1] *Institut für Theoretische Physik, Goethe Universität, Max-von-Laue-Str. 1, 60438, Frankfurt am Main, Germany*
[2] *Mullard Space Science Laboratory, University College London, Holmbury St. Mary, Dorking, Surrey, RH5 6NT, UK*
[3] *Max-Planck-Institut für Radioastronomie, Auf dem Hügel 69, Bonn 53121, Germany*
[4] *Department of Astrophysics/IMAPP, Radboud University Nijmegen P.O. Box 9010, 6500 GL Nijmegen, The Netherlands*
[5] *Jodrell Bank Centre for Astrophysics, University of Manchester, M13 9PL, UK.*
[6] *Frankfurt Institute for Advanced Studies, Ruth-Moufang-Str. 1, Frankfurt am Main, 60438, Germany*

**Parametrised generic metric for a spherically symmetric black hole**

As mentioned in the Letter, although an analytic form for the metric of a general dilaton-axion black hole (BH) is known, our calculations have been performed using a general parameterisation of BH spacetimes in arbitrary metric theories of gravity. This approach was first developed for spherically symmetric spacetimes by Rezzolla and Zhidenko[1] and it has been more recently extended to axisymmetric spacetimes[2]. The advantage of this approach is that it provides a general and agnostic framework within which to perform the analysis of accretion flows onto BHs, the shadow of arbitrary BHs[3], or the dynamics of test objects on orbits close to the BH[4]. For example, the deviation from Einstein's general relativity can be expressed using a small number of coefficients.

In the case considered here of a non-rotating dilaton BH spacetime, the approach uses a continued-fraction expansion (Padé expansion) along the radial coordinate to express the two non-trivial metric functions. More specifically, the approach starts by taking the line element of any spherically symmetric BH to be expressed (adopting the Lorentzian signature) as

$$\mathrm{d}s^2 = -N(r)\,\mathrm{d}t^2 + \frac{B(r)}{N(r)}\,\mathrm{d}r^2 + r^2\,\mathrm{d}\theta^2 + r^2\sin^2\theta\,\mathrm{d}\phi^2\,, \qquad (1)$$

where $N(r)$ and $B(r)$ are functions of the radial coordinate, $r$. The position of event horizon is fixed at $r = r_0 > 0$, which implies that $N(r_0) = 0$ by construction. Furthermore, the radial coordinate is compactified through a dimensionless variable

$$x := 1 - \frac{r_0}{r}, \qquad (2)$$

where $x = 0$ corresponds to the position of the event horizon and $x = 1$ to spatial infinity. Furthermore, the metric function $N(r)$ is written as

$$N(x) = x\,A(x), \qquad (3)$$

where $A(x) > 0$ for $0 \leq x \leq 1$. Introducing additional coefficients $\epsilon$, $a_n$, and $b_n$ (where $n$ is the expansion order), the metric functions $A(r)$ and $B(r)$ are expressed in terms of $x$ as follows

$$A(x) = 1 - \epsilon(1-x) + (a_0 - \epsilon)(1-x)^2 + \widetilde{A}(x)(1-x)^3, \qquad (4)$$
$$B(x) = 1 + b_0(1-x) + \widetilde{B}(x)(1-x)^2, \qquad (5)$$

where $\widetilde{A}$ and $\widetilde{B}$ are functions which simultaneously describe the metric near the event horizon and at spatial infinity. In particular, $\widetilde{A}$ and $\widetilde{B}$ have rapid convergence properties due to the Padé expansion and are expressed as follows

$$\widetilde{A}(x) = \cfrac{a_1}{1 + \cfrac{a_2 x}{1 + \cfrac{a_3 x}{1 + \ldots}}}, \qquad \widetilde{B}(x) = \cfrac{b_1}{1 + \cfrac{b_2 x}{1 + \cfrac{b_3 x}{1 + \ldots}}}, \qquad (6)$$

where $a_1, a_2, a_3 \ldots$ and $b_1, b_2, b_3 \ldots$ are dimensionless coefficients that can, in principle, be constrained by observations. The dimensionless parameter $\epsilon$ is fixed by both $r_0$ and the ADM mass $M$, measuring the deviation from the Schwarzschild case as

$$\epsilon := \frac{2M - r_0}{r_0} = -\left(1 - \frac{2M}{r_0}\right). \qquad (7)$$

Any static spacetime can be directly simulated in the GRMHD code BHAC[5] and GRRT images may be constructed using BHOSS (Younsi et al. 2018 in prep.) [3]. In practice, within the aforementioned parameterisation, expansion to at least second order ($n = 2$) is required to maintain sufficient numerical accuracy in all metric terms, enabling GRMHD simulations of the chosen spacetime.

**Parameterisation for a Non-rotating Dilaton BH**

We next use the parameterisation described in the previous section to express the metric of a generic asymptotically flat dilaton-axion BH spacetime[6] in the non-rotating limit. This is obtained after reducing the axially symmetric solution of Einstein-Maxwell-dilaton-axion gravity theory, which is the low energy limit of the bosonic sector of the heterotic string theory[7,8]. In this theory, the Einstein-Maxwell equations are coupled with a scalar field, i.e., the dilaton field.

When both the axion field and the BH spin vanish, a dilaton-axion BH is described by a spherically symmetric metric with line element

$$\mathrm{d}s^2 = -\left(\frac{r-2\mu}{r+2\hat{b}}\right)\mathrm{d}t^2 + \left(\frac{r+2\hat{b}}{r-2\mu}\right)\mathrm{d}\rho^2 + (r^2+2\hat{b}r)\,\mathrm{d}\Omega^2\,. \tag{8}$$

The pseudo-radial coordinate $\rho$ and the ADM mass $M$ are expressed as

$$\rho^2 = r^2 + 2\hat{b}r\,, \qquad M = \mu + \hat{b}\,, \tag{9}$$

where $\hat{b}$ is the dilaton parameter, which has an allowed range of $0 \leq \hat{b} \leq M$ [9]. The coefficients of the dilaton BH[1] are now given by

$$\epsilon = \sqrt{1+\frac{\hat{b}}{\mu}} - 1\,, \tag{10}$$

$$a_0 = \frac{\hat{b}}{2\mu}\,, \tag{11}$$

$$b_0 = 0\,, \tag{12}$$

$$a_1 = 2\sqrt{1+\frac{\hat{b}}{\mu}} + \frac{1}{1+\hat{b}/(2\mu)} - 3 - \frac{\hat{b}}{2\mu}\,, \tag{13}$$

$$b_1 = \frac{\sqrt{1+\hat{b}/\mu}}{1+\hat{b}/(2\mu)} - 1\,, \tag{14}$$

$$a_2 = \frac{\sqrt{1+\hat{b}/\mu} - \frac{1}{2} - \hat{b}^2/(4\mu^2) + \hat{b}/(2\mu)\left(\sqrt{1+\hat{b}/\mu}-1\right)}{\left(1+\hat{b}/(2\mu)\right)^2}\,, \tag{15}$$

$$b_2 = \frac{\sqrt{1+\hat{b}/\mu}}{1+\hat{b}/(2\mu)} - 1 - \frac{\hat{b}^2}{(\hat{b}+2\mu)^2}\,. \tag{16}$$

If $\hat{b}=0$, the coefficients $a_0$, $a_1$ and $b_1$ become zero (Eqs. (4) and (5) vanish by construction), and the line element recovers the Schwarzschild metric. Even in the case of $\hat{b}=0.5\,M$, the maximum relative difference between the exact and the expanded expressions for the metric function $g_{tt}$ (at second order) is $\sim 3\times 10^{-4}$, as shown by Rezzolla & Zhidenko[1].

**Location of characteristic radii for Kerr BH and dilaton BH**

Although the Kerr and dilaton BH solutions are mathematically and physically distinct from one another, it is still possible to make a sensible comparison between these two spacetimes when certain characteristic radii are matched between the two solutions. The three key characteristic radii we match are: (i) the event horizon, (ii) the (unstable) photon orbit, and (iii) the innermost stable circular orbit (ISCO). Further details may be found in the main Letter. For the non-rotating dilaton BH, the radius of the event horizon is given by

$$r_{\text{h,dilaton}} = 2\left(M - \hat{b}\right), \tag{17}$$

whereas for a Kerr BH (with spin parameter $a := J/M$) it is given by

$$r_{\text{h,Kerr}} = M + \sqrt{M^2 - a^2}. \tag{18}$$

Although. frame-dragging is an important property of the Kerr metric that is absent in the spherically symmetric dilaton BH solution considered here, it is interesting to determine the values of the dilaton parameter which yield a BH having, on the equatorial plane and for corotating photons, the same event horizon, unstable photon orbit, or ISCO radius of a Kerr BH of given angular momentum. To this end equations (17) and (18) are equated, yielding the value of the dilaton parameter as a function of $a$ as

$$\hat{b}_{\text{h}} = \frac{1}{2}\left(M - \sqrt{M^2 - a^2}\right). \tag{19}$$

For a given choice of the Kerr BH spin parameter, eq. (19) will therefore provide the value of the dilaton parameter $\hat{b}_{\text{h}}$ whose corresponding dilaton BH will have an event horizon radius equal to that of the Kerr BH (again, in the equatorial plane).

Similarly, the ISCO for particles circulating in the equatorial plane may be determined by setting to zero the effective potential, along with its first and second derivatives, and solving for $r$. For a spherically-symmetric spacetime this yields[10]

$$E^2 \frac{d^2 g_{\phi\phi}}{dr^2} + L_z^2 \frac{d^2 g_{tt}}{dr^2} + \frac{d^2}{dr^2}\left(g_{tt} g_{\phi\phi}\right) = 0, \tag{20}$$

where the particle's energy, $E$, and angular momentum, $L_z$, are given respectively by

$$E := -u^t g_{tt}, \tag{21}$$
$$L_z := \Omega u^t g_{\phi\phi}, \tag{22}$$

and where $\Omega$ (angular velocity) and $u^t$ are then given by

$$\Omega := \frac{u^\phi}{u^t} = \left(-\frac{dg_{tt}}{dr} \bigg/ \frac{dg_{\phi\phi}}{dr}\right)^{1/2}, \tag{23}$$
$$u^t = \left(-g_{tt} - \Omega^2 g_{\phi\phi}\right)^{-1/2}. \tag{24}$$

Solving Eq. (20) with Eqs. (21)–(24) yields the ISCO radius of the dilation BH as

$$r_{\text{ISCO}} = 2M \left( \mathcal{B} + \mathcal{B}^2 + \mathcal{B}^3 \right), \tag{25}$$

where $\mathcal{B}$ is defined as

$$\mathcal{B} := \left( 1 - \frac{\hat{b}}{M} \right)^{1/3}. \tag{26}$$

Similar to the derivation of (19), equating the dilaton ISCO radius and the Kerr ISCO radius ($r_{\text{K,ISCO}}$, see Bardeen et al. 1972[11]), the dilaton parameter as a function of $a$ is obtained as

$$\hat{b}_{\text{ISCO}} = M \left[ 1 + \frac{1}{27} \left( 1 + \sigma - \frac{2}{\sigma} \right)^3 \right], \tag{27}$$

where $\sigma$ is defined as

$$\sigma^3 := \frac{-14M + 3 \left( -9\, r_{\text{K,ISCO}} + \sqrt{36M^2 + 84M\, r_{\text{K,ISCO}} + 81\, r_{\text{K,ISCO}}^2} \right)}{4M}. \tag{28}$$

Finally, the radius of the (unstable) photon orbit may be calculated from Eq. (24) as

$$r_{\text{photon}} = \frac{1}{2} \left[ 3(M-b) + \sqrt{(M-b)(9M-b)} \right], \tag{29}$$

from which upon equating with the expression for the Kerr photon orbit radius yields the dilaton parameter expressed in terms of $a$ as

$$\hat{b}_{\text{photon}} = \frac{1}{2} M \left( -2 - 3\mathcal{C} + \sqrt{8 + \mathcal{C}(\mathcal{C}+8)} \right), \tag{30}$$

where $\mathcal{C}$ is defined as

$$\mathcal{C} := \cos \left[ \frac{2}{3} \cos^{-1} \left( -\frac{a}{M} \right) \right]. \tag{31}$$

Recalling that in the Letter the Kerr spin parameter is specified to be $a = 0.6\, M$, which gives $r_{\text{K,ISCO}} = 3.829\, M$, the corresponding values of the dilaton parameter for which the Kerr BH and dilaton BH event horizon, photon orbit, and ISCO radii coincide are $\hat{b} = 0.1\, M$, $0.339\, M$, and $0.504\, M$, respectively.

**Mass accretion and magnetic flux rates onto the BH**

In order to investigate more quantitatively the properties of the accretion flows onto the two different BHs we calculate both the volume-integrated mass accretion rate and the magnetic flux rate at the BH horizon, showing them in Figure 1. In particular, we define the mass accretion rate as

$$\dot{M} := \int_0^{2\pi} \int_0^{\pi} \rho u^r \sqrt{-g}\, d\theta\, d\phi, \tag{32}$$

while the magnetic flux rate is given by

$$\dot{\Phi}_B := \frac{1}{2} \int_0^{2\pi} \int_0^{\pi} |B^r| \sqrt{-g} \, d\theta \, d\phi \,. \qquad (33)$$

The mass accretion and magnetic flux rates exhibit very similar behaviour for both cases. Initially a large amount of mass accretes onto the BH, bringing with it a large magnetic flux. The mass accretion rate then gradually decreases with time until it reaches a quasi-steady state at around $t = 6000 \, M$, when the magnetorotational instability (MRI) is fully developed. Analogous behaviour is also observed in the time evolution of the magnetic flux rate.

**Possible April 2017 EHTC configuration**

For the April 2017 EHTC campaign we assume that the total observing time is shared between different target sources and that a total time span of 6.0 hours (including pointing and focussing time) is available for observations of Sgr A*. In Figure 3 we show the *UV*-coverage and resulting visibility amplitude for different 6 hour time slots during the 7th of April 2017. The early time slot from UTC 03:00–09:00 shows a very sparse sampling of the *UV*-plane in comparison to the full *UV*-coverage of the EHTC (see panels a and b in Supplementary Figure 3). The two later time slots show similar *UV*-coverage (panels **c** and **d**) and visibility amplitudes (panels **g** and **h**). However, the UTC 08:30–14:30 time slot is favoured due to the longer measurement period of the Hawaiian telescopes.

**Closure Phases**

Closure phases, e.g., the sum of the phases of a closed antenna triangle, are most sensitive to the physical parameters of the Galactic Center. Thus they can be used to study the source asymmetry and variability[12,13]. The variation of the closure phase for two selected triangles (JCMT-SPT-LMT and SMA-SPT-ALMA) is presented in Figure 5. Both triangles show non-zero closure phases which indicate an asymmetric source structure.

**Alternative MEM algorithms**

To test how the reconstruction algorithm affects our results, we perform an image comparison between BSMEM[14] and PolMEM[15]. The results of this test are presented in Figure 6 and Supplementary Table 3, while for the Kerr image and VLBI data we use the values presented in Tables 1 and 2. Overall, this test shows that both reconstruction algorithms provide very similar results and in general perform very well, with a slightly better DSSIM value for BSMEM. This supports our choice of the reconstruction algorithm used throughout this work.

**Alternative comparisons: matched horizons and photon orbits**

Here we present comparisons similar to those shown in Fig. 3 of the Letter, however now matching the two BHs such that they have the same event horizon (Figure 7, middle row) or the same (unstable) photon orbit (Figure 7, bottom row). Note that the same considerations made when matching the ISCO radii also apply here and underline the robustness of the conclusions drawn in this work. Similar to the ISCO case, we compute the MSE and DSSIM values (see Supplementary Table 4). Due to the larger southern limb in the horizon matched case, the centroid of the reconstructed images is shifted more to the South. In contrast to the bright southern limb, the structure of the dim northern limb is not clearly visible in the reconstructed image (see middle row in Figure 7). A better matching between the GRRT convolved and reconstructed images is obtained in the case where the photon orbits are matched. The slight offset of the centroid is caused by the larger spread of the flux density in the left arc. Besides this small deviation, the images are in good agreement (see Table 4). In both cases, interstellar scattering enhances the blurring of the features.

**Future developments and improvements**

As mentioned in the main Letter, there are several future developments that can increase our ability to distinguish between shadow images from GR and from alternative theories of gravity. In what follows we present two possible improvements to the EHT and the impact they have: for simplicity we restrict our analysis to a Kerr BH and a dilaton BH with matched ISCOs.

As a first test, we consider the impact of adding two antennas in the southern region of the African continent. The first telescope is placed at Gamsberg (Namibia) and second in Drakensberg (South Africa). For both antennas we assume a design similar to the APEX telescope, noting that the inclusion of the African telescopes extends the *UV* sampling in E-W direction and improves the sampling on intermediate *UV*-scales ($-4\,G\lambda < (U, V) < 4\,G\lambda$). In order to benefit from the extended array, the observing time was increased to 12 hours, observing from UTC 03:00–15:00. The GRRT images are therefore averaged between $11000\,M$ and $13000\,M$.

The results of the synthetic imaging using BSMEM are presented in Figure 8. The top row corresponds to the Kerr BH and the bottom row to the dilaton BH. In this particular case the images are convolved with $50\%$ of the EHT beam size, plotted in the lower left corner of each panel. Panels **a** and **d** show the convolved GRRT images and the synthetic images based on the 2017 EHT configurations are presented in panels **b** and **e**, while the impact of the EHT configuration extended with the inclusion of two African antennas (denoted as EHT2017 + Africa) can be seen in panels **c** and **f**. Clearly the addition of the two African telescopes and the increased observing time lead to a improved reconstruction, as indicated by the extent and location of emission centroid in the left limb in both images. This behaviour is also reflected in the decreased DSSIM values (from $0.314$ to $0.135$ in the case of the Kerr BH and from $0.401$ to $0.158$ in the dilaton case).

As a second test, we consider the effect of increasing the observing frequency from $230\,\text{GHz}$ to $340\,\text{GHz}$ and the bandwidth from $4\,\text{GHz}$ to $16\,\text{GHz}$, as this may also improve the image quality and thus the capability of the array to test different theories of gravity. Besides increasing the observing frequency we included the aforementioned two additional stations in southern Africa. The increase of the observing frequency is accompanied by a rise in the SEFD of each antenna, which leads to a higher thermal noise.

In Figure 9 we present the synthetic $340\,\text{GHz}$ images for both the Kerr and the dilaton BHs, while in Table 5 we list the $340\,\text{GHz}$ SEFD for each antenna used in our synthetic imaging. The GGRT images convolved with $50\%$ of the EHT beam are presented in panels **a** and **c** and the reconstructed images as seen by the extended EHT including two African antennas (panels **b** and **d**). The reconstructed images at $340\,\text{GHz}$ agree very well with the GRRT images and show very detailed features, e.g., the splitting of the southern limb into two regions in the case of the dilaton BH (see panels **c** and **d** in Figure 8).

In summary, although our ability to distinguish BH spacetimes from shadow images alone is somewhat limited at the present time, the results presented here clearly indicate that technological developments will improve this ability, motivating further work in this direction. We note that a Bayesian analysis would be an alternative approach for searching the parameter range of Sgr A*, e.g., observer inclination angle, black-hole spin, position angle, etc. Essential pre-requisites for such an analysis, however, are simple but accurate semi-analytic models of radiatively inefficient accretion flow with which to fit the observed visibilities; while some attempts have been made in this respect[16,17], a more systematic exploration of the space of parameters via fully nonlinear numerical simulations is needed.

**Supplementary Information References**

Supplementary Table 1: **Chosen antenna parameters.** Shown are the antenna name, the effective antenna diameter $d$, and the system equivalent flux density (SEFD).

| Telescope | $d_{\text{eff}}$ [m] | SEFD [Jy] |
|---|---|---|
| ALMA | 85 | 110 |
| APEX | 12 | 6500 |
| JCMT | 15 | 4700 |
| LMT | 50 | 560 |
| PDB | 37 | 5200 |
| PV | 30 | 2900 |
| SMT | 10 | 11900 |
| SMA | 23 | 4900 |
| SPT | 10 | 1600 |

Supplementary Table 2: **Chosen observation parameters.**

| Parameter | Value |
|---|---|
| scan length | 420 s |
| integration time | 12 s |
| off-source time | 600 s |
| start time | 2017:097:08:30:00 (UT) |
| end time | 2017:097:14:30:00 (UT) |
| bandwidth | 4096 MHz |

Supplementary Table 3: **Image statistics: the mean square error (MSE) and the structural dissimilarity (DSSIM) index between Image 1 and Image 2.** Italic letters correspond to the various panels in Figure 6. Small values of the MSE or of the DSSIM represent very well-matched images.

| Image 1 | Image 2 | MSE$_{1,2}$ | DSSIM$_{1,2}$ |
|---|---|---|---|
| Kerr $a$ | Kerr $b$ (PolMEM) | 0.015 | 0.22 |
| Kerr $a$ | Kerr $c$ (BSMEM) | 0.012 | 0.21 |

Supplementary Table 4: **Image statistics: the mean square error (MSE) and the structural dissimilarity (DSSIM) index between Image 1 and Image 2.** Italic letters correspond to the various panels in Fig. 7. Small values of the MSE or of the DSSIM represent very well-matched images.

| Image 1 | Image 2 | MSE$_{1,2}$ | DSSIM$_{1,2}$ |
|---|---|---|---|
| Kerr $a$ | Kerr $b$ [Kerr $c$] | 0.016 [0.016] | 0.31 [0.31] |
| Dilaton $d$ | Dilaton $e$ [Dilaton $f$] | 0.011 [0.011] | 0.44 [0.45] |
| Dilaton $g$ | Dilaton $h$ [Dilaton $i$] | 0.016 [0.016] | 0.49 [0.51] |

Supplementary Table 5: **Chosen antenna parameters.** Shown are the antenna name, the effective antenna diameter $d$, and the system equivalent flux density (SEFD) at 340 GHz.

| Telescope | $d_{\text{eff}}$ [m] | SEFD [Jy] |
|---|---|---|
| ALMA | 85 | 140 |
| APEX | 12 | 12200 |
| JCMT | 15 | 9400 |
| LMT | 50 | 13700 |
| PDB | 37 | 3400 |
| PV | 30 | 5200 |
| SMT | 10 | 23100 |
| SMA | 23 | 8100 |
| SPT | 10 | 3400 |
| GAMS | 12 | 12200 |
| DRAK | 12 | 12200 |

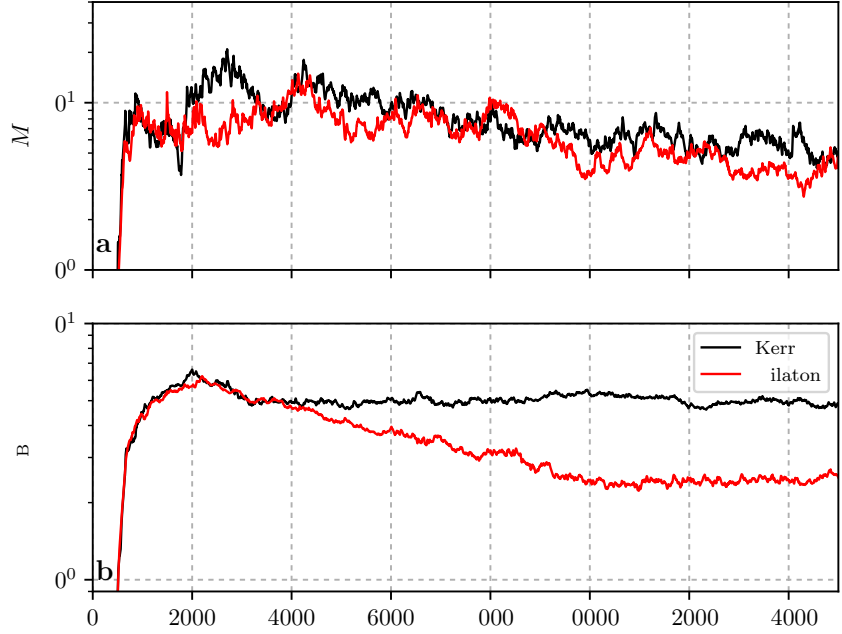

Supplementary Figure 1: **Evolution of the mass accretion rate and of the magnetic flux rate.** Mass accretion rate (panel **a**) and magnetic-flux rate (panel **b**) as calculated at the event horizon of an $a = 0.6\,M$ Kerr BH (black lines) and of a non-rotating dilaton BH with $\hat{b} = 0.504\,M$ (red lines). Both BHs are matched to have the same ISCO radius.

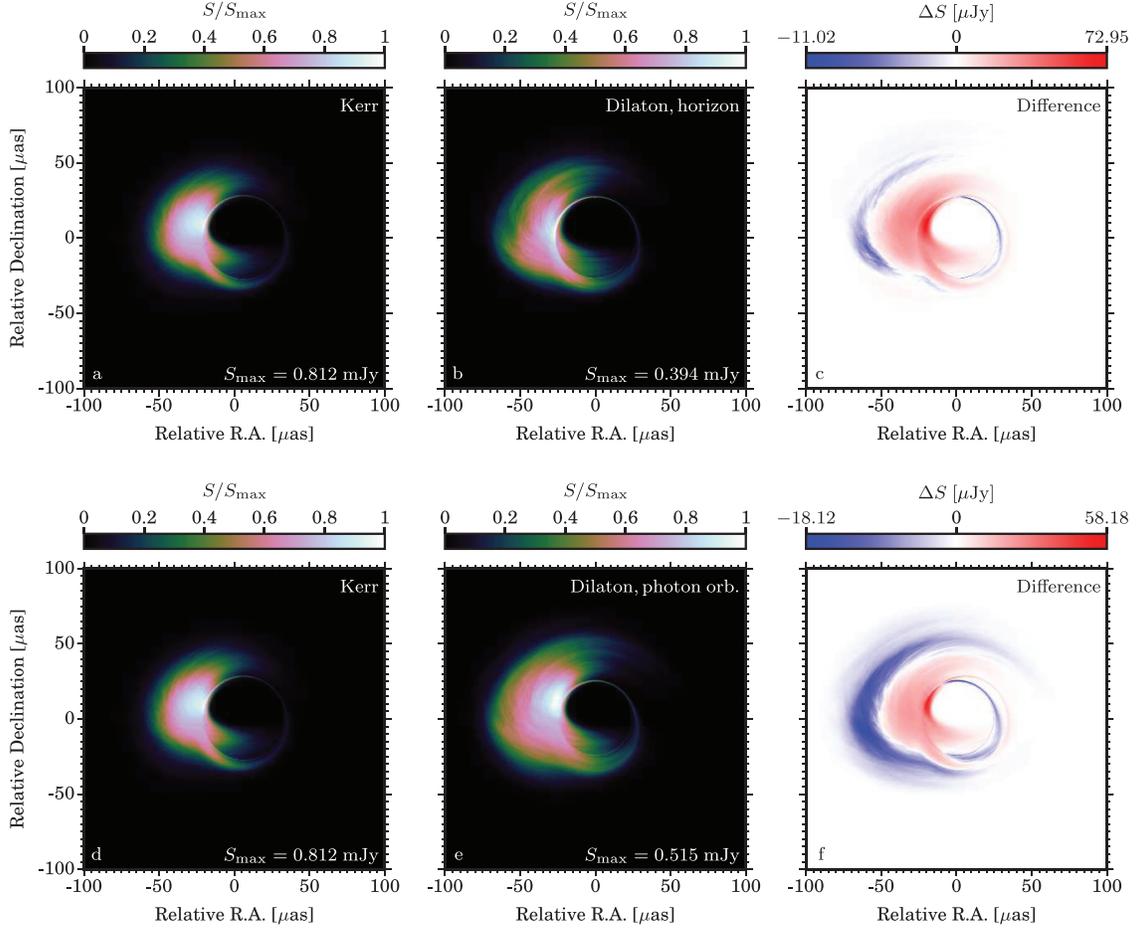

Supplementary Figure 2: **Simulated BH shadow images of Sgr A* calculated from GRMHD simulations of an accreting BH.** Five-and-a-half-hour-averaged simulated BH shadow images of Sgr A* of GRMHD simulations of accretion flow: onto a Kerr BH (panels **a** and **d**), onto a non-rotating dilaton BH matched to the event horizon radius of the Kerr BH (panel **b**), and onto a non-rotating dilaton BH matched with the (unstable) photon orbit radius of the Kerr BH (panel **e**). Panels **c** and **f** show a pixel-by-pixel image difference comparison of the Kerr BH case with respect to the dilaton BH cases. The colour scale is linear and uniformly increasing in intensity. In the pixel-by-pixel difference image, red denotes pixels from the Kerr BH image being brighter than the dilaton BH image, blue pixels denote the converse, and near-white indicates very close agreement between both images.

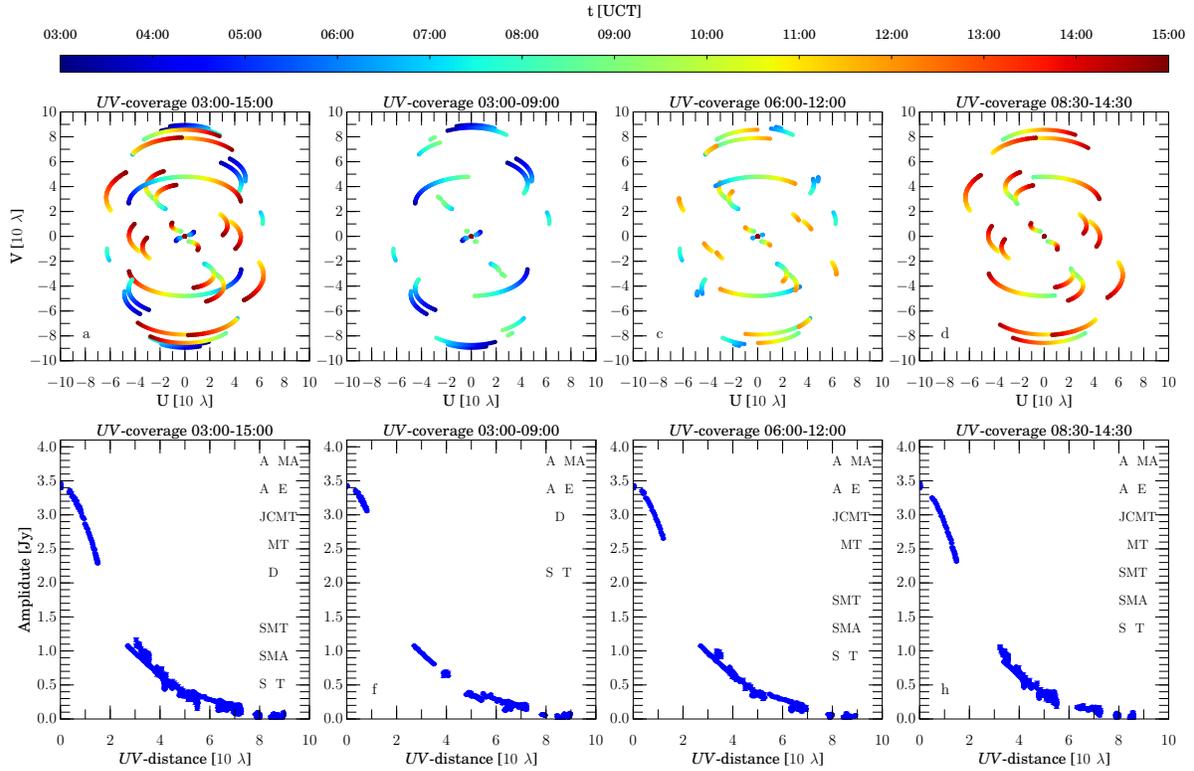

Supplementary Figure 3: **UV-plane and visibility amplitude for different EHTC configurations.** Top row, from left to right: *UV*-plane with temporal variation colour-coded (panels **a**–**d**). Bottom row: visibility amplitudes for the above different time slots (12 hours for panel **e** and 6.0 hours for panels **f**–**h**). The participating antennas are listed on the right in each panel.

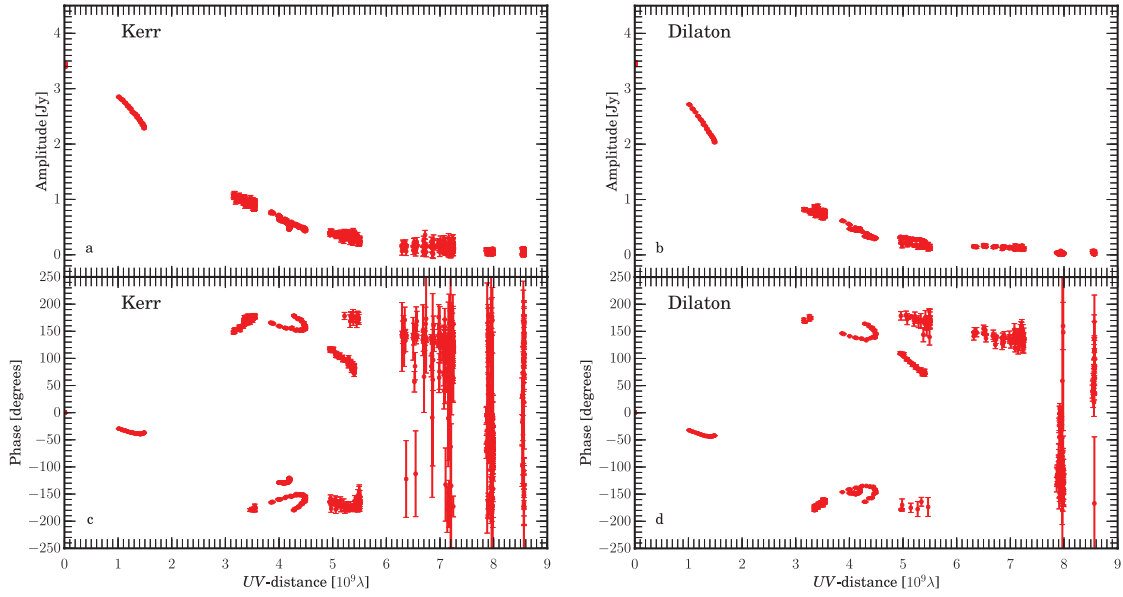

Supplementary Figure 4: **Visibility amplitude and visibility phase.** Upper panels show visibility amplitude (panels **a** and **b**) and lower panels present visibility phase as function of the baseline length (panels **c** and **d**) for the Kerr BH case (*left*) and non-rotating dilaton BH case (*right*). In this visibility data, thermal noise and 10% random errors on the telescope gains are included. For visual clarity the data is downsampled such that only every tenth data point is plotted.

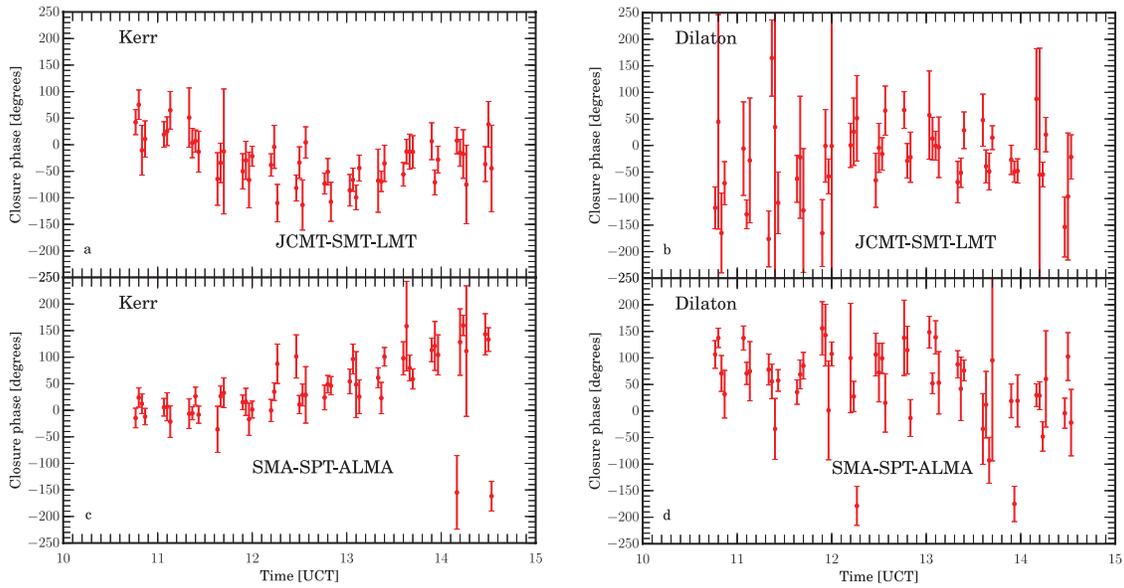

Supplementary Figure 5: **Variation of the closure phase.** Panels show the variation of the closure phases of two different triangles, *upper:* JCMT-SPT-LMT (panels **a** and **b**), *lower:* SMA-SPT-ALMA, (panels **c** and **d**) for the Kerr BH case (*left*) and non-rotating dilaton BH case (*right*). For visual clarity the data is downsampled such that only every tenth data point is plotted.

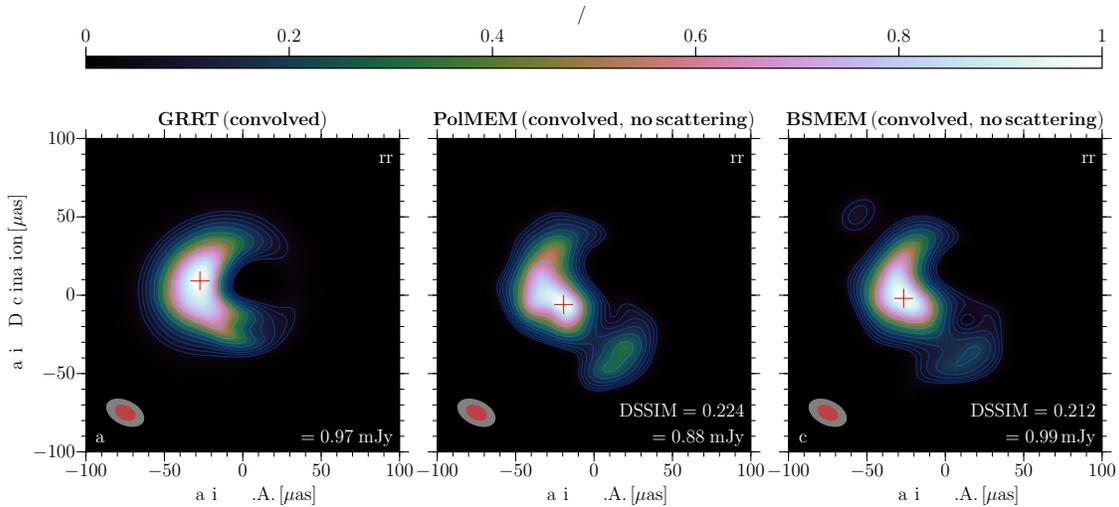

Supplementary Figure 6: **Synthetic BH shadow images of Sgr A* for the Kerr BH using two different image re-construction algorithms.** From left to right: GRRT image convolved with $50\%$ (red shading) of the nominal beam size (light grey shading), the contour levels start at $5\%$ of the peak value and increase by $\sqrt{2}$. The red cross in the images marks the position of the flux density maximum (panel **a**), reconstructed image without interstellar scattering using PolMEM convolved with $50\%$ red shading) of the nominal beam size (light grey shading, panel **b**), and reconstructed image without interstellar scattering using BSMEM (panel **c**). All images are based on visibilities which take into account a possible VLBI antenna configuration and schedule for the EHTC April 2017 observations. The convolving beam is plotted in the lower left corner of each panel (see *SI* for details).

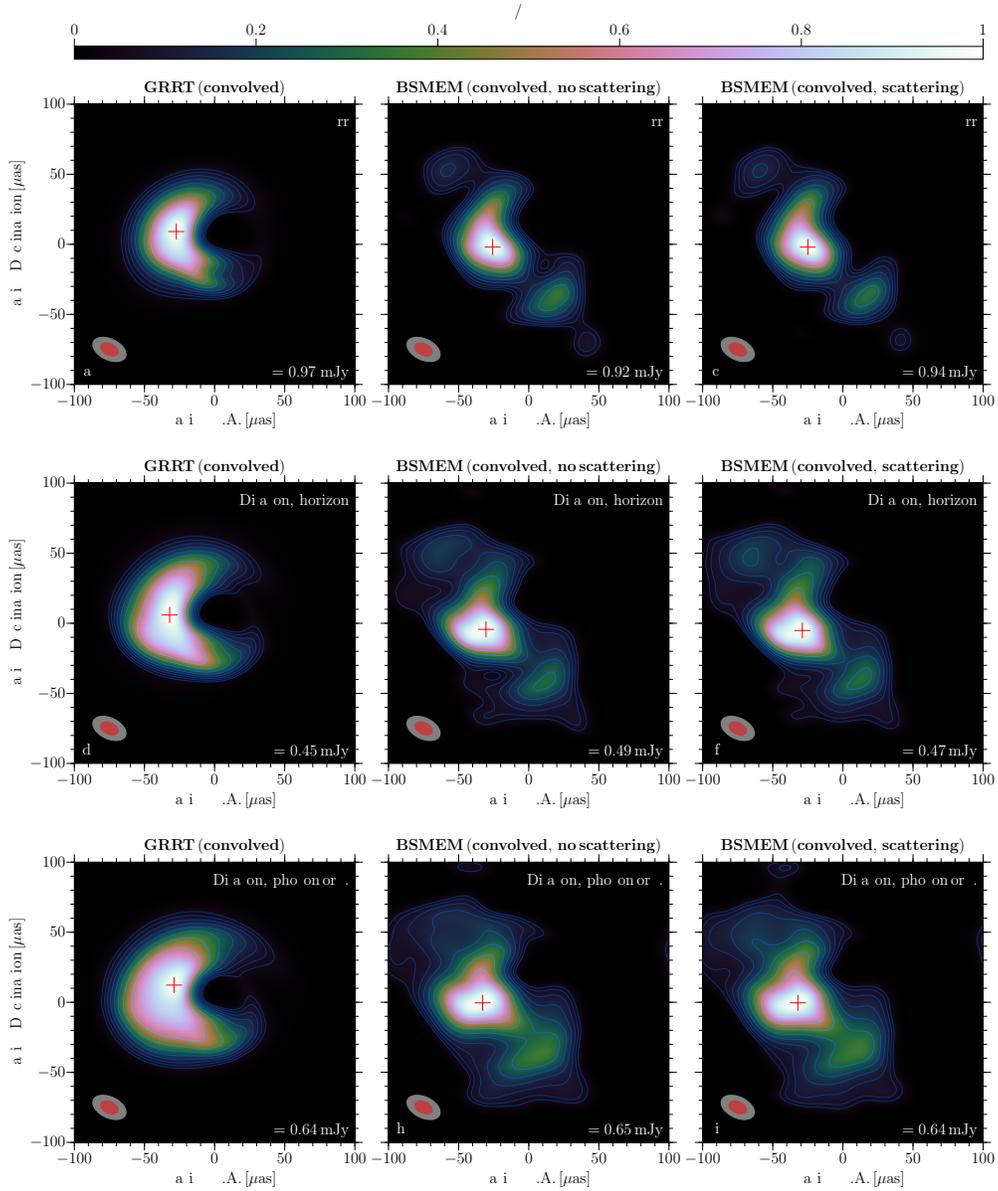

Supplementary Figure 7: **Synthetic BH shadow images of Sgr A\* for a Kerr BH (*panels a–c*), for non-rotating dilaton BH with matched horizon (*panels d–f*), and for a non-rotating dilaton BH with matched photon orbit (*panels g–i*).** From left to right and for all rows: GRRT image convolved with 50% (red shading) of the nominal beam size (light grey shading), the contour levels start at 5% of the peak value and increase by $\sqrt{2}$. The red cross in the images marks the position of the flux density maximum (panels **a**, **d**, **g**); reconstructed image without interstellar scattering using BSMEM convolved with 50% (red shading) of the nominal beam size (light grey shading, panels **b**, **e**, **h**); and reconstructed image including the effect of interstellar scattering using BSMEM (panels **c**, **f**, **i**). All images are based on visibilities which take into account a possible VLBI antenna configuration and schedule for the EHTC April 2017 observations.

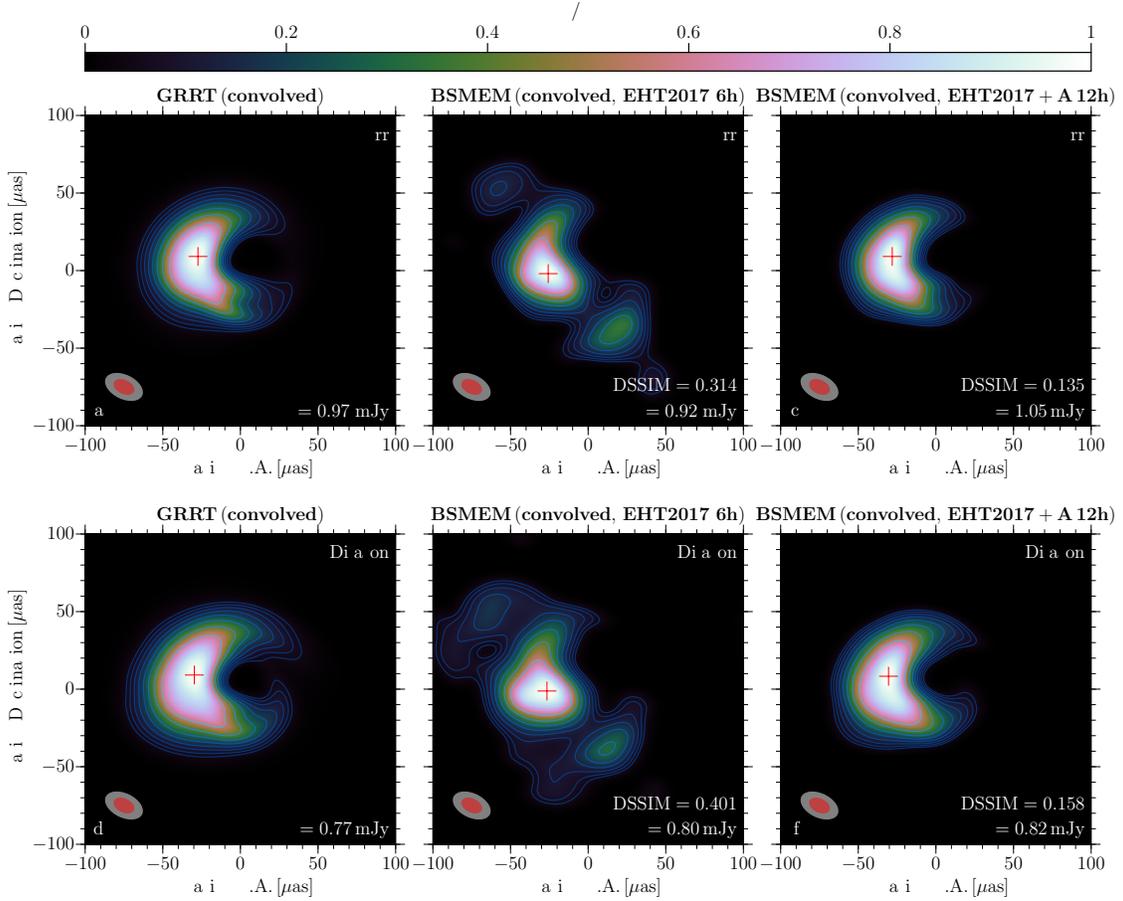

Supplementary Figure 8: **Synthetic BH shadow images of Sgr A* for a Kerr BH (*panels a–c*) and for non-rotating dilaton BH with matched ISCO (*panels d–f*).** From left to right and for all rows: GRRT image convolved with 50% (red shading) of the nominal beam size (light grey shading), the contour levels start at 5% of the peak value and increase by $\sqrt{2}$. The red cross in the images marks the position of the flux density maximum (panels **a** and **d**); reconstructed image without interstellar scattering using BSMEM convolved with 50% (red shading) of the nominal beam size (light grey shading) using a possible EHTC April 2017 configuration (panels **b** and **e**); and reconstructed image without interstellar scattering using BSMEM convolved with 50% (red shading) of the nominal beam size (light grey shading) using a possible future EHT configuration (panels **c** and **f**). The convolving beam is plotted in the lower left corner of each panel.

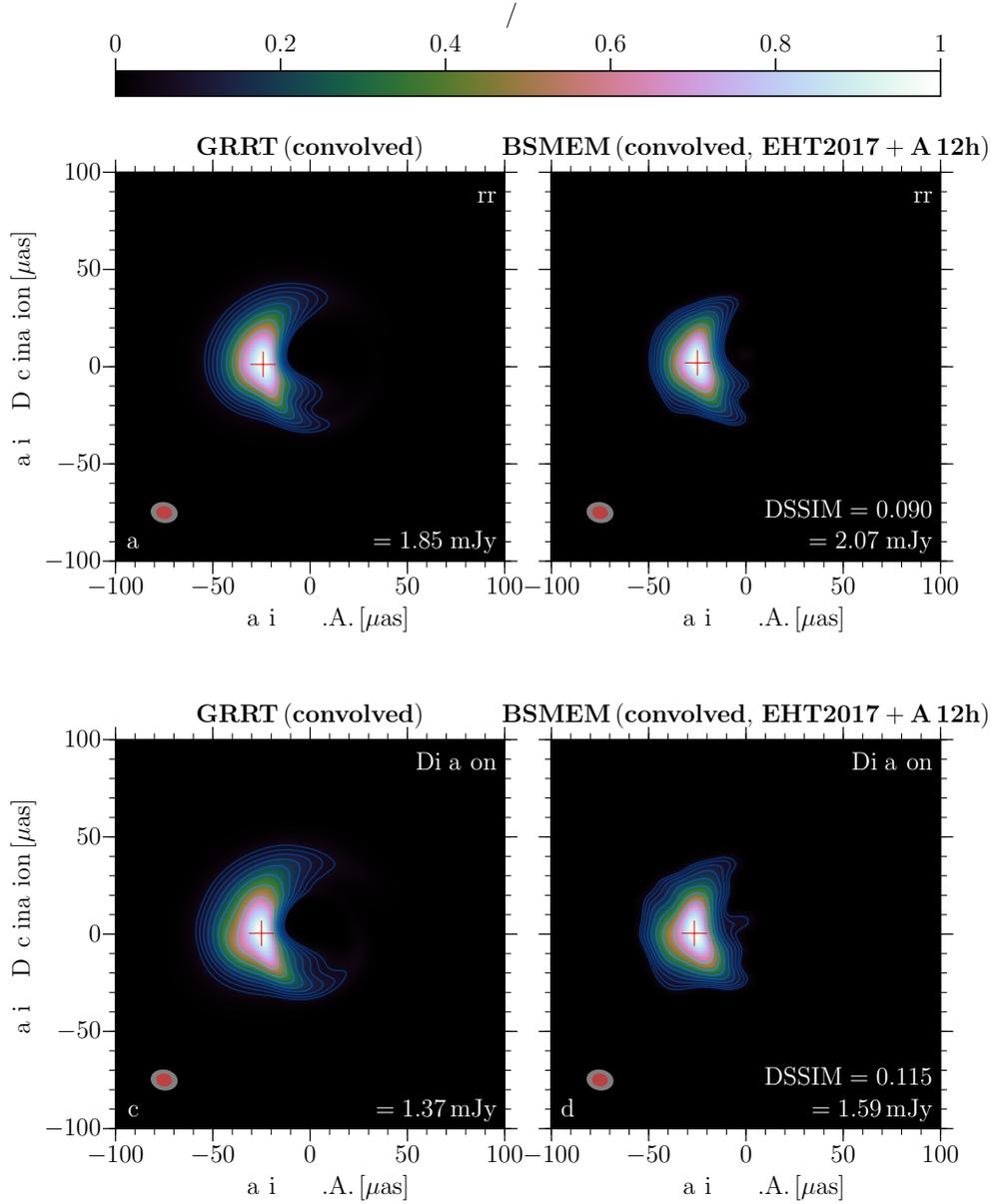

Supplementary Figure 9: **340 GHz Synthetic BH shadow images of Sgr A* for a Kerr BH (*panels a and b*) and for non-rotating dilaton BH with matched ISCO (*panels c and d*).** From left to right and for all rows: GRRT image convolved with 50% (red shading) of the nominal beam size (light grey shading), the contour levels start at 5% of the peak value and increase by $\sqrt{2}$. The red cross in the images marks the position of the flux density maximum (panels **a** and **c**); reconstructed image without interstellar scattering using BSMEM convolved with 50% (red shading) of the nominal beam size (light grey shading) using a possible future EHT configuration (panels **b** and **d**). The convolving beam is plotted in the lower left corner of each panel.

21